\documentclass[prl,aps,twocolumn,superscriptaddress,showpacs]{revtex4}

\usepackage{longtable}
\usepackage{graphicx}
\usepackage{dcolumn}

\def\ifmath#1{\relax\ifmmode #1\else $#1$\fi}%
\def\GeV{\ifmmode \hbox{\rm\ GeV}\else
                  \hbox{GeV}\fi}%
\def\MeV{\ifmmode \hbox{\rm\ MeV}\else
                  \hbox{MeV}\fi}%
\def\keV{\ifmmode \hbox{\rm\ keV}\else
                  \hbox{keV}\fi}%
\def\eV{\ifmmode \hbox{\rm\ eV}\else
                 \hbox{eV}\fi}%

\newcommand{\et}{\ifmath{E_T}}

\newcommand{\Ht}{\ifmath{H_T}}

\def\z0{\ifmath{Z^0}}%
\def\Z0{\ifmath{Z^0}}

\newcommand{\ttbar}{\ifmath{t\overline{t}}}
\newcommand{\ppbar}{\ifmath{p\overline{p}}}
\newcommand{\qqbar}{\ifmath{q\overline{q}}}
\newcommand{\bbbar}{\ifmath{b\overline{b}}}
\newcommand{\ccbar}{\ifmath{c\overline{c}}}
\newcommand{\pt}{\ifmath{p_T}}

\newcommand{\met}{\ifmath {\not\!\!E_T}}

\newcommand{\secvtx}{{\tt SecVtx}}
\newcommand{\pythia}{{\sc pythia}}
\newcommand{\herwig}{{\sc herwig}}
\newcommand{\alpgen}{{\sc alpgen}}
\newcommand{\geant}{{\sc geant}}
\newcommand{\madevent}{{\sc madevent}}
\newcommand{\qq}{{\sc qq}}

\newcommand{\wbb}{\ifmath{Wb\overline{b}}}
\newcommand{\wcc}{\ifmath{Wc\overline{c}}}
\newcommand{\wc}{\ifmath{Wc}}
\newcommand{\sigmatt}{\ifmath{\sigma_{\ttbar}}}

\newcommand{\crossresult}[5]{\ifmath{#1^{+#2}_{-#3} (\rm {stat.}) ^{+#4}_{-#5} (\rm {syst.})~\rm{pb} }}
\newcommand{\crossresultssa}[5]{\ifmath{#1 \pm {#2} (\rm {stat.}) ^{+#4}_{-#5} (\rm {syst.})~\rm{pb} }}

\newcommand{\VPY}[3]{{\bf #1}, #2 (#3)}
\newcommand{\PRD}[3]{Phys. Rev. D \VPY{#1}{#2}{#3}}
\newcommand{\PRL}[3]{Phys. Rev. Lett. \VPY{#1}{#2}{#3}}
\newcommand{\PLB}[3]{Phys. Lett. B \VPY{#1}{#2}{#3}}
\newcommand{\NP}[3]{Nucl. Phys. \VPY{#1}{#2}{#3}}
\newcommand{\NIM}[3]{Nucl. Instrum. Methods A \VPY{#1}{#2}{#3}}
\newcommand{\JHEP}[3]{J. High Energy Phys. \VPY{#1}{#2}{#3}}

\def\etal{\emph{et al.}}

\def\xsL{8.7}
\def\xsLStatU{0.9}
\def\xsLStatD{0.9}
\def\xsLSystU{1.1}
\def\xsLSystD{0.9}

\def\xsLD{10.1}
\def\xsLDStatU{1.6}
\def\xsLDStatD{1.4}
\def\xsLDSystU{2.0}
\def\xsLDSystD{1.3}

\def\xsT{8.7}
\def\xsTStatU{0.9}
\def\xsTStatD{0.9}
\def\xsTSystU{1.1}
\def\xsTSystD{0.9}

\def\xsJP{8.9}
\def\xsJPStatU{1.0}
\def\xsJPStatD{1.0}
\def\xsJPSystU{1.1}
\def\xsJPSystD{1.0}

\def\xs{\xsL}
\def\xsStatU{\xsLStatU}
\def\xsStatD{\xsLStatD}
\def\xsSystU{\xsLSystU}
\def\xsSystD{\xsLSystD}

\def\xsD{\xsLD}
\def\xsDStatU{\xsLDStatU}
\def\xsDStatD{\xsLDStatD}
\def\xsDSystU{\xsLDSystU}
\def\xsDSystD{\xsLDSystD}

\begin{document}


\title{\Large\bf Measurement of the \boldmath $t\overline{t}$ Production Cross Section 
	in $p\overline{p}$ Collisions at $\sqrt{s} = 1.96$~TeV} 

\affiliation{Institute of Physics, Academia Sinica, Taipei, Taiwan 11529, Republic of China} 
\affiliation{Argonne National Laboratory, Argonne, Illinois 60439} 
\affiliation{Institut de Fisica d'Altes Energies, Universitat Autonoma de Barcelona, E-08193, Bellaterra (Barcelona), Spain} 
\affiliation{Baylor University, Waco, Texas  76798} 
\affiliation{Istituto Nazionale di Fisica Nucleare, University of Bologna, I-40127 Bologna, Italy} 
\affiliation{Brandeis University, Waltham, Massachusetts 02254} 
\affiliation{University of California, Davis, Davis, California  95616} 
\affiliation{University of California, Los Angeles, Los Angeles, California  90024} 
\affiliation{University of California, San Diego, La Jolla, California  92093} 
\affiliation{University of California, Santa Barbara, Santa Barbara, California 93106} 
\affiliation{Instituto de Fisica de Cantabria, CSIC-University of Cantabria, 39005 Santander, Spain} 
\affiliation{Carnegie Mellon University, Pittsburgh, PA  15213} 
\affiliation{Enrico Fermi Institute, University of Chicago, Chicago, Illinois 60637} 
\affiliation{Joint Institute for Nuclear Research, RU-141980 Dubna, Russia} 
\affiliation{Duke University, Durham, North Carolina  27708} 
\affiliation{Fermi National Accelerator Laboratory, Batavia, Illinois 60510} 
\affiliation{University of Florida, Gainesville, Florida  32611} 
\affiliation{Laboratori Nazionali di Frascati, Istituto Nazionale di Fisica Nucleare, I-00044 Frascati, Italy} 
\affiliation{University of Geneva, CH-1211 Geneva 4, Switzerland} 
\affiliation{Glasgow University, Glasgow G12 8QQ, United Kingdom} 
\affiliation{Harvard University, Cambridge, Massachusetts 02138} 
\affiliation{Division of High Energy Physics, Department of Physics, University of Helsinki and Helsinki Institute of Physics, FIN-00014, Helsinki, Finland} 
\affiliation{University of Illinois, Urbana, Illinois 61801} 
\affiliation{The Johns Hopkins University, Baltimore, Maryland 21218} 
\affiliation{Institut f\"{u}r Experimentelle Kernphysik, Universit\"{a}t Karlsruhe, 76128 Karlsruhe, Germany} 
\affiliation{High Energy Accelerator Research Organization (KEK), Tsukuba, Ibaraki 305, Japan} 
\affiliation{Center for High Energy Physics: Kyungpook National University, Taegu 702-701, Korea; Seoul National University, Seoul 151-742, Korea; and SungKyunKwan University, Suwon 440-746, Korea} 
\affiliation{Ernest Orlando Lawrence Berkeley National Laboratory, Berkeley, California 94720} 
\affiliation{University of Liverpool, Liverpool L69 7ZE, United Kingdom} 
\affiliation{University College London, London WC1E 6BT, United Kingdom} 
\affiliation{Centro de Investigaciones Energeticas Medioambientales y Tecnologicas, E-28040 Madrid, Spain} 
\affiliation{Massachusetts Institute of Technology, Cambridge, Massachusetts  02139} 
\affiliation{Institute of Particle Physics: McGill University, Montr\'{e}al, Canada H3A~2T8; and University of Toronto, Toronto, Canada M5S~1A7} 
\affiliation{University of Michigan, Ann Arbor, Michigan 48109} 
\affiliation{Michigan State University, East Lansing, Michigan  48824} 
\affiliation{Institution for Theoretical and Experimental Physics, ITEP, Moscow 117259, Russia} 
\affiliation{University of New Mexico, Albuquerque, New Mexico 87131} 
\affiliation{Northwestern University, Evanston, Illinois  60208} 
\affiliation{The Ohio State University, Columbus, Ohio  43210} 
\affiliation{Okayama University, Okayama 700-8530, Japan} 
\affiliation{Osaka City University, Osaka 588, Japan} 
\affiliation{University of Oxford, Oxford OX1 3RH, United Kingdom} 
\affiliation{University of Padova, Istituto Nazionale di Fisica Nucleare, Sezione di Padova-Trento, I-35131 Padova, Italy} 
\affiliation{LPNHE, Universite Pierre et Marie Curie/IN2P3-CNRS, UMR7585, Paris, F-75252 France} 
\affiliation{University of Pennsylvania, Philadelphia, Pennsylvania 19104} 
\affiliation{Istituto Nazionale di Fisica Nucleare Pisa, Universities of Pisa, Siena and Scuola Normale Superiore, I-56127 Pisa, Italy} 
\affiliation{University of Pittsburgh, Pittsburgh, Pennsylvania 15260} 
\affiliation{Purdue University, West Lafayette, Indiana 47907} 
\affiliation{University of Rochester, Rochester, New York 14627} 
\affiliation{The Rockefeller University, New York, New York 10021} 
\affiliation{Istituto Nazionale di Fisica Nucleare, Sezione di Roma 1, University of Rome ``La Sapienza," I-00185 Roma, Italy} 
\affiliation{Rutgers University, Piscataway, New Jersey 08855} 
\affiliation{Texas A\&M University, College Station, Texas 77843} 
\affiliation{Istituto Nazionale di Fisica Nucleare, University of Trieste/\ Udine, Italy} 
\affiliation{University of Tsukuba, Tsukuba, Ibaraki 305, Japan} 
\affiliation{Tufts University, Medford, Massachusetts 02155} 
\affiliation{Waseda University, Tokyo 169, Japan} 
\affiliation{Wayne State University, Detroit, Michigan  48201} 
\affiliation{University of Wisconsin, Madison, Wisconsin 53706} 
\affiliation{Yale University, New Haven, Connecticut 06520} 
\author{A.~Abulencia}
\affiliation{University of Illinois, Urbana, Illinois 61801}
\author{D.~Acosta}
\affiliation{University of Florida, Gainesville, Florida  32611}
\author{J.~Adelman}
\affiliation{Enrico Fermi Institute, University of Chicago, Chicago, Illinois 60637}
\author{T.~Affolder}
\affiliation{University of California, Santa Barbara, Santa Barbara, California 93106}
\author{T.~Akimoto}
\affiliation{University of Tsukuba, Tsukuba, Ibaraki 305, Japan}
\author{M.G.~Albrow}
\affiliation{Fermi National Accelerator Laboratory, Batavia, Illinois 60510}
\author{D.~Ambrose}
\affiliation{Fermi National Accelerator Laboratory, Batavia, Illinois 60510}
\author{S.~Amerio}
\affiliation{University of Padova, Istituto Nazionale di Fisica Nucleare, Sezione di Padova-Trento, I-35131 Padova, Italy}
\author{D.~Amidei}
\affiliation{University of Michigan, Ann Arbor, Michigan 48109}
\author{A.~Anastassov}
\affiliation{Rutgers University, Piscataway, New Jersey 08855}
% ...
\collaboration{CDF Collaboration}
\noaffiliation

\date{\today}

\begin{abstract}

\hspace{\parindent}
 
We present a measurement of the top quark pair production cross section in 
\ppbar\ collisions at $\sqrt{s}=1.96$ TeV using 318 pb$^{-1}$ of data 
collected with the Collider Detector at Fermilab.  
We select \ttbar\ decays into the final states $e\nu$ +~jets and
$\mu\nu$ +~jets, in which at least one $b$ quark from the $t$-quark decays is identified
using a secondary vertex-finding algorithm.  
Assuming a top quark mass of 178 GeV/$c^2$, 
we measure a cross section of \crossresultssa{\xs}{\xsStatU}{\xsStatD}{\xsSystU}{\xsSystD}.  
We also report the first observation of \ttbar\ with significance greater than
$5 \sigma$  
in the subsample in which both $b$ quarks are identified,
corresponding to a cross section of 
\crossresult{\xsD}{\xsDStatU}{\xsDStatD}{\xsDSystU}{\xsDSystD}.

\end{abstract}



\maketitle


The top quark completes the third quark generation in the Standard Model
(SM).  Due to its large mass, nearly 40 times greater than that of the next
heaviest quark, the top quark can at present only be studied at the Fermilab Tevatron,
a \ppbar\ collider with a center-of-mass energy ($\sqrt{s}$) of 1.96 TeV.  
In these collisions, the SM top quark is mostly produced in pairs through \qqbar\ 
annihilation and gluon fusion with a total theoretical 
cross section of $6.1_{-0.8}^{+0.6}$~pb~\cite{ref:theory_xs} 
for a mass of 178~GeV/$c^2$~\cite{ref:run1mass}.  
Previous measurements of the \ttbar\ production cross section (\sigmatt) performed
at the Tevatron~\cite{ref:run2_xs,ref:run2_xs_other}
were consistent with SM expectations, 
but suffered from large uncertainties due to small event samples.
With significant enhancements in the integrated
luminosity and our sensitivity to $t$-quark decay products, we are no
longer limited by statistical uncertainties, now reaching a precision comparable
to that of the theory.


The top quark decays predominantly to a $W$ boson and a bottom quark. 
We expect the top quark signal significance to be greatest in the lepton+jets channel,
in which top quark pairs decay through the chain $t\overline{t} \to
W^+W^-b\overline{b} \to \ell\nu q\overline{q}' b\overline{b}$.
A typical \ttbar\ event in this decay channel will include an electron 
or muon with large transverse momentum ($p_T$), four jets corresponding
to the four final-state quarks, and an 
imbalance in the total transverse energy in the event (\met) from 
the undetected neutrino~\cite{ref:coords}.
We distinguish \ttbar\ events from background by requiring at least one jet
to be identified as a bottom quark ($b$-tagged) using the secondary
vertex-finding algorithm (\secvtx) described in Ref.~\cite{ref:run2_xs}.  
We have re-optimized the algorithm to enhance the efficiency for 
$b$-tagging two jets per event; 
the sample of events with this signature
is dominated by \ttbar\ ($> 90\%$ pure), making it an ideal environment for
direct determination of top quark properties.  
In this Letter, we present a measurement of the \ttbar\ cross section (\sigmatt) in
$b$-tagged lepton+jets events, and we also report the first observation 
of \ttbar\ with significance greater than $5 \sigma$ in doubly $b$-tagged events.


Results reported here are obtained
using  318~pb$^{-1}$ of integrated luminosity collected 
between March 2002 and August 2004 by
the Collider Detector at Fermilab (CDF II).
The CDF II detector~\cite{ref:CDF_detector} is a general-purpose particle detector located at one of
the two interaction points at the Tevatron Collider. 
Inside a 1.4~T solenoidal magnetic field, a large open-cell drift chamber, 
the Central Outer Tracker (COT)~\cite{ref:cot}, and an eight-layer silicon 
system~\cite{ref:silicon} provide tracking information. 
The COT covers the pseudo-rapidity range $|\eta| < 1.1$ and
provides a long lever arm for track curvature measurements.  
The silicon system provides three-dimensional
hit information between radii of 1.3~cm and 28~cm, with
an $r-\phi$ impact parameter resolution of $\sim 40 \mu$m (including a 30~$\mu$m 
contribution from the beamspot).
Outside the solenoid, electromagnetic and hadronic 
calorimeters surround the tracking volume in a projective tower geometry, identifying jets and electrons
with $|\eta| < 3.6$.  
Electron energies are measured in the central electromagnetic calorimeter for
$|\eta| < 1.1$ and in the end-plug calorimeters for $1.1 < |\eta| < 3.6$.  
Beyond the calorimeters, drift chambers provide muon identification in the region $|\eta| < 1.0$.


The data were collected with an inclusive high-\pt\ lepton trigger that requires
an electron (muon) with transverse energy $\et > 18$ GeV  ($\pt > 18$ GeV/$c$).
The trigger efficiency, including lepton identification, 
is 95.9 $\pm$ 1.5\% (87.7 $\pm$ 1.5\%) for electrons (muons). 
Following the event selection used in Ref.~\cite{ref:run2_xs},
we require all leptons to be isolated~\cite{ref:isolation} and
$\et > 20$~GeV ($\pt > 20$~GeV/$c$) for electrons (muons).
We remove events with multiple high-\pt\ leptons, a cosmic ray muon, a photon conversion electron, 
or a track that forms the $Z$ mass with the lepton.  
The position of the primary vertex along the beam is 
required to be within 60 cm of the nominal interaction position and consistent 
with the $z$ position of the high-\pt\ lepton.

After leptons are selected, we require the presence of at least three 
jets with $|\eta|<2$ and $\et > 15 \GeV$ (\et\ is corrected as described 
in Ref.~\cite{ref:jetcorr}).  
Jets are clustered with a cone-based algorithm with a cone size 
$\Delta R \equiv \sqrt{\Delta \phi^2 + \Delta \eta^2} =0.4$.

To account for the expected neutrino, we require large missing transverse energy, 
$\met > 20$~GeV.  
The invariant mass of the lepton and the neutrino, with the
longitudinal neutrino momentum set to zero, is required to exceed 
20~GeV/$c^2$; this rejects $\sim 50\%$ of background events that do not contain 
a real $W$ boson.
Finally, as \ttbar\ events typically have larger total transverse energy than 
background events, we require that the scalar sum of the \met\ and
the lepton and total jet transverse energies (\Ht) be greater than 200~GeV.  
This leaves 310 (190) events in the electron (muon) data sample,
dominated by $W$ bosons with associated production of light-flavor jets ($W$+ LF).  
Table I includes the event count before $b$-tagging (pretag) sorted by the number of 
jets in the event.  
The one- and two-jet bins have been included as an assumed control sample,
although without the \Ht\ requirement.

%
%
We improve the \ttbar\ signal significance by requiring at least one jet to be 
$b$-tagged with the \secvtx\ algorithm~\cite{ref:run2_xs}.
Due to their long lifetime, $b$ quarks typically decay a measurable distance
from the primary interaction point.
We reconstruct the decay vertices using a minimum of two or three tracks with 
an impact parameter significance 
greater than 3.0 or 2.0, respectively.  
Contributions from $K^0_S$ and $\Lambda$ decays and interactions in the inner detector 
material are reduced through additional requirements on the invariant mass of the 
secondary vertex and an upper limit on the track impact parameter, $d_0 < 0.15$~cm.
We measure the two-dimensional displacement of the secondary vertex 
from the primary interaction point projected along the jet axis ($L_{2D}$);
a jet is $b$-tagged if the vertex has $L_{2D}$ significance larger than 6.0,
where the uncertainty on $L_{2D}$ includes contributions from both the primary 
and secondary vertex fits.
The probability of misidentifying a light-flavor jet as a $b$-quark jet due to 
detector resolution (mistag rate) is estimated from secondary vertices reconstructed 
behind the primary vertex with $L_{2D}$ significance less than -6.0.


The \ttbar\ acceptance is calculated from a combination of data and
Monte Carlo simulation. 
We use the \pythia\ Monte Carlo generator~\cite{ref:pythia} with CTEQ5L 
parton distribution functions~\cite{ref:cteq5l} and 
assume $M_{top} = 178$ GeV/$c^2$, the world average top quark mass measurement from 
Run I~\cite{ref:run1mass}.
The CLEO \qq\ Monte Carlo program~\cite{ref:cleomc} models the decays 
of bottom and charm hadrons. 
These events are passed through a \geant~\cite{ref:geant} simulation of the
CDF II detector and subjected to the same selection requirements as the data.
The total acceptance, including the branching fraction, is calculated as the 
product of geometric and kinematic acceptances (including lepton 
identification) and the trigger efficiency.  
The efficiency to identify isolated high-\pt\ leptons in the simulation
is scaled to the value measured in $Z\to\ell^+\ell^-$ data.  
The acceptance for electron (muon) events before $b$-tagging
is $3.8 \pm 0.3\%$ ($2.8 \pm 0.2\%$).

The $b$-tagging efficiency for the full \ttbar\ event is measured 
in a \ttbar\ simulation that has been tuned to
match the single jet $b$-tagging efficiency in the data.
A multiplicative scale factor ($S_b = 0.927 \pm 0.066$), measured in
heavy-flavor-enriched samples of non-isolated, low-\et\ leptons, 
corrects for the per-jet efficiency difference 
between data and simulation~\cite{ref:run2_xs}.
In \ttbar\ events, we measure a $b$-tagging efficiency of $48\pm4\%$ 
for $b$-quark jets with a corresponding mistag rate of $1.2 \pm 0.1\%$; 
we therefore expect $69 \pm 5\%$ ($23\pm3\%$) of these events to 
have at least one $b$-tag (two $b$-tags).


The systematic uncertainty in the cross section due to the $S_b$ 
correction is 6\%, 
dominated by the extrapolation from the charm-contaminated low-\et\ sample to 
more energetic \ttbar\ events.  
Other leading systematic uncertainties arise from the integrated luminosity 
measurement (6\%)~\cite{ref:lum}, jet energy corrections (5\%), 
lepton isolation (2\%), lepton identification (2\%), 
parton distribution functions (2\%), choice of Monte Carlo generator 
(2\%), and modeling of initial and final state radiation (1\%).
These systematic uncertainties are uncorrelated for \ttbar\ and are added in 
quadrature for the signal expectation.


The background to \ttbar\ is mostly due to direct production of a $W$ boson
with multiple jets ($W$+jets).  
Smaller contributions come from QCD jet production in which the $W$ signature 
is faked by jets appearing as electrons or by semi-leptonic $b$-hadron 
decays (non-$W$), and electroweak processes such as single top quark production, 
diboson ($WW$, $WZ$, and $ZZ$) production, and $Z$ boson decays to tau pairs. 
We describe these backgrounds in turn and summarize the results in 
Tables~\ref{tab:singletag} and~\ref{tab:doubletag}.
The total uncertainty accounts for correlations between the individual 
background sources and the \ttbar\ prediction where appropriate.

We separate the contribution from $W$+jets into events with and without 
heavy-flavor jets.  
For the former, the fractions of $W$+jets events attributable to 
\wbb, \wcc, and \wc\ are estimated with \alpgen/\herwig\ 
Monte Carlo~\cite{ref:alpgen,ref:herwig}, then scaled by a multiplicative factor 
of $1.5 \pm 0.4$ to reproduce the $b$-tag rates observed in a control 
sample of inclusive jet data. 
The expected number of events is estimated by multiplying these fractions by the number
of pretag events, after removing the pretag expectations for all other backgrounds 
and \ttbar\ signal.  

%
%
We estimate the background contribution from $W$ events with only light-flavor 
jets by applying the mistag rate, measured in the inclusive 
jet dataset and 
parameterized in jet $E_T$, $\eta$, $\phi$, and number of tracks and the total 
jet energy in the event, to the lepton+jets dataset.  
The mistag rate is adjusted higher by $36 \pm 13\%$ to account 
for heavy-flavor contamination in the jet data and residual contributions 
from material interactions and $K^0_S$/$\Lambda$ decays.  
Finally, this result is adjusted down by the fraction of the data sample 
attributed to physics processes with heavy-flavor production.

%
%
The expectation for the non-$W$ background is determined using data.
Assuming that the isolation of the lepton and the \met\ 
in these events are uncorrelated~\cite{ref:run2_xs}, 
we extrapolate from the low-\met\ and 
non-isolated regions (which contain fewer real $W$ bosons) to 
predict the non-$W$ content of the pretag and signal sample.  

Finally, we use Monte Carlo to estimate the backgrounds due to 
single top (\pythia/\madevent~\cite{ref:madevent}) and dibosons/$Z\to\tau\tau$ (\pythia), 
normalizing the expectations to their respective theoretical 
cross sections~\cite{ref:ew-theory}.
Here, the $b$-tagging efficiency is evaluated analogously to the \ttbar\ 
signal prediction.

\begin{center}
\begin{figure}[htb]
\includegraphics[width=2.9in]{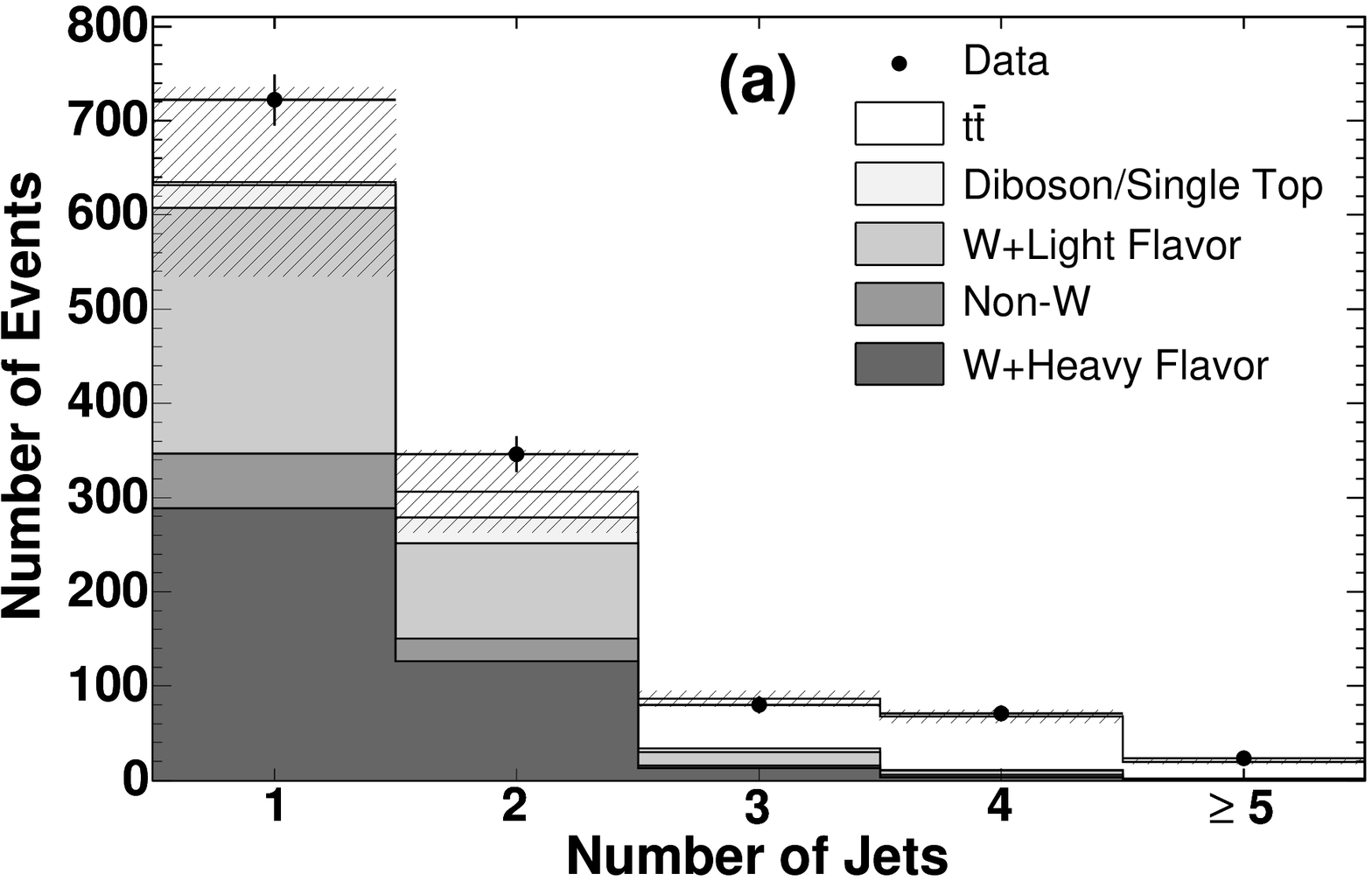}
\includegraphics[width=2.9in]{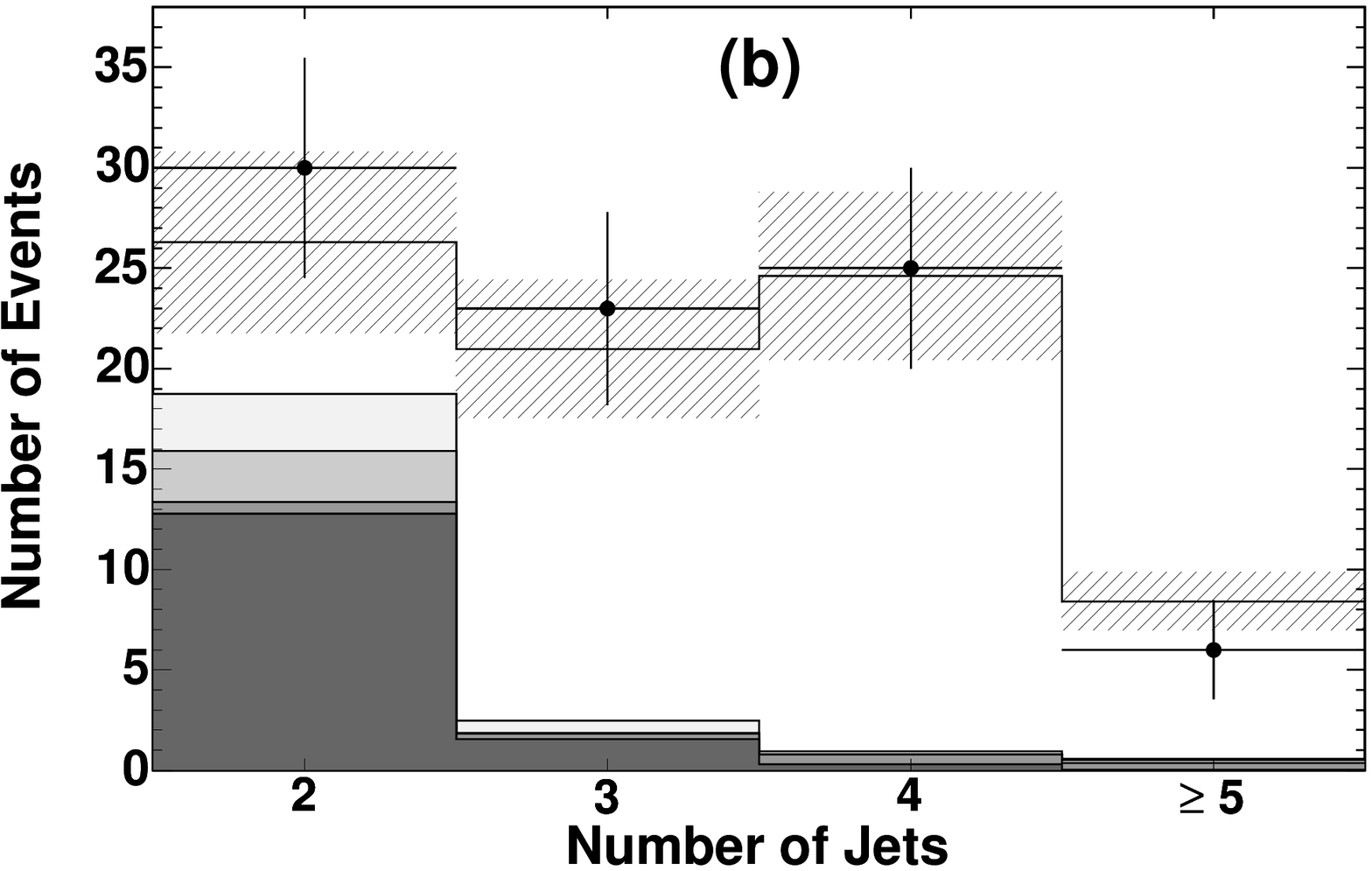}
    \caption{Summary of background and signal event yields versus number of 
	jets in the event when requiring (a) at least one $b$-tagged jet  
	and (b) at least two $b$-tagged jets.	
	The \ttbar\ contribution is normalized to the measured cross section in each 
	sample.
	The $H_T$ requirement is released for events with fewer than 3~jets.
	The hashed region shows the uncertainty on the total expectation.}
    \label{fig:singletag} 
\end{figure}
\end{center}

Backgrounds that depend on the assumed \ttbar\ cross section are calculated 
iteratively.  
The signal and background contributions to the $b$-tagged data sample,
sorted by jet multiplicity,
are summarized in Table~\ref{tab:singletag} and Figure~\ref{fig:singletag}(a).
The total corrected background in the signal region with at least one $b$-tag 
is $46 \pm 9$ events, where we observe 174 events.  
We interpret the excess of events with three or more jets as pure \ttbar\ signal, 
corresponding to a cross section of 
$\crossresultssa{\xs}{\xsStatU}{\xsStatD}{\xsSystU}{\xsSystD}$.

We observe 54 events with multiple $b$-tags, 
the first time the event yield is inconsistent with the no-top hypothesis 
with significance greater than $5\sigma$.  
After correcting for \ttbar, we expect $4.1 \pm 2.5$ background events, 
a significant improvement in signal purity (Table~\ref{tab:tagsummary}). 
We summarize these results in 
Table~\ref{tab:doubletag} and Figure~\ref{fig:singletag}(b);
the agreement in five-jet events underscores our ability to model initial- 
and final-state gluon radiation.
We measure a cross section of $\crossresult{\xsD}{\xsDStatU}{\xsDStatD}{\xsDSystU}{\xsDSystD}$
in the multiply $b$-tagged sample, 
consistent with the result for events with a single $b$-tag.  

The acceptance and efficiency both have a small dependence on the top quark mass; 
the cross section measurements change by $\pm 0.08$~pb for each $\mp 1$~GeV/$c^2$ 
change in the assumed top quark mass from the initial value of 178~GeV/$c^2$, 
in the range of 160 -- 190~GeV/$c^2$.

We perform two cross checks using alternate 
$b$-tagging algorithms.
Table~\ref{tab:tagsummary} compares the $b$-tagging characteristics of 
these checks with those of the main analyses.
In the first, we repeat the analysis with an update of the original \secvtx\ 
algorithm described in Ref.~\cite{ref:run2_xs} ({\tt T\secvtx}).  
We observe 138 events with at least one $b$-tag, 
over an expected background of $25 \pm 5$ events.  
The measured cross section is 
$\crossresultssa{\xsT}{\xsTStatU}{\xsTStatD}{\xsTSystU}{\xsTSystD}$.
In the second cross check, we use the {\tt Jet Probability} 
algorithm~\cite{JetProb}. 
We observe 120 events with at least one $b$-tag compared to a 
background of $21 \pm 3$ events, corresponding to 
$\sigma_{t\bar{t}} = \crossresultssa{\xsJP}{\xsJPStatU}{\xsJPStatD}{\xsJPSystU}{\xsJPSystD}$.
Both cross checks are in agreement with the lead result above, 
albeit with highly correlated uncertainties.

In summary, we have measured a \ttbar\ production cross section of 
$\crossresultssa{\xs}{\xsStatU}{\xsStatD}{\xsSystU}{\xsSystD}$
in \ppbar\ 
collisions at $\sqrt{s}=1.96$~TeV using 318~pb$^{-1}$ of data with 
at least one secondary vertex $b$-tag.  
The result is consistent with the SM expectation of $6.1_{-0.8}^{+0.6}$~pb
for a top quark mass of 178~GeV/$c^2$, 
which changes by $\pm 0.2$ pb for every $\mp 1$~GeV/$c^2$ 
shift in the assumed mass. 
Additionally, we have studied a large, very pure sample of \ttbar\ events with 
at least two $b$-tags, which will form the foundation for future high-precision 
measurements of top quark properties.  In this sample, we measure 
$\sigmatt = \crossresult{\xsD}{\xsDStatU}{\xsDStatD}{\xsDSystU}{\xsDSystD}$.

We thank the Fermilab staff and the technical staffs of the participating institutions for their vital contributions. This work was supported by the U.S. Department of Energy and National Science Foundation; the Italian Istituto Nazionale di Fisica Nucleare; the Ministry of Education, Culture, Sports, Science and Technology of Japan; the Natural Sciences and Engineering Research Council of Canada; the National Science Council of the Republic of China; the Swiss National Science Foundation; the A.P. Sloan Foundation; the Bundesministerium f\"ur Bildung und Forschung, Germany; the Korean Science and Engineering Foundation and the Korean Research Foundation; the Particle Physics and Astronomy Research Council and the Royal Society, UK; the Russian Foundation for Basic Research; the Comisi\'on Interministerial de Ciencia y Tecnolog\'{\i}a, Spain; in part by the European Community's Human Potential Programme under contract HPRN-CT-2002-00292; and the Academy of Finland.


\begingroup
\begin{table*}[hbt]
\caption{
Summary of event yields and background (Bkgd) expectations 
sorted by the number of jets in the event.
Event totals before $b$-tagging (Pretag) are listed in the first row; 
all other entries correspond to the sample with at least one $b$-tagged jet,
assuming the measured \ttbar\ cross section of 8.7~pb.
The $H_T$ requirement is released for events with fewer than 3~jets. }

\label{tab:singletag}
\begin{ruledtabular}
\begin{tabular}{lccccc}
	& $W + 1$ jet	& $W + 2$ jets	& $W + 3$ jets	& $W + 4$ jets	& $W + \geq$ 5 jets\\
\hline
Pretag	& 30,283	& 4676	& 324	& 142	& 34 \\
\hline
Dibosons & 6.3 $\pm$ 0.9 & 11.8 $\pm$ 1.7 & 1.7 $\pm$ 0.3 & 0.46 $\pm$ 0.12 & 0.15 $\pm$ 0.04\\
$t$ 	& 8.1 $\pm$ 2.9	& 13.2 $\pm$ 4.0  & 1.8 $\pm$ 0.5 & 0.39 $\pm$ 0.12 & 0.08 $\pm$ 0.03 \\
$Z\to\tau\tau$	& 9.7 $\pm$ 1.9	& 2.5 $\pm$ 0.5	& 0.18 $\pm$ 0.04 & 0.017 $\pm$ 0.003	& 0.008 $\pm$ 0.002 \\
$W\bbbar$	& 114 $\pm$ 35	& 63 $\pm$ 19		& 6.1 $\pm$ 1.6		& 1.4 $\pm$ 0.7		& 0.17 $\pm$ 0.09 \\
$W\ccbar$	& 46 $\pm$ 13	& 30 $\pm$ 9		& 3.8 $\pm$ 1.2		& 1.1 $\pm$ 0.4		& 0.13 $\pm$ 0.04 \\
$Wc$	        & 128 $\pm$ 33	& 34 $\pm$ 9		& 3.0 $\pm$ 0.8		& 0.81 $\pm$ 0.21	& 0.10 $\pm$ 0.03 \\
$W + LF$	& 261 $\pm$ 57	& 101 $\pm$ 22	        & 14.4 $\pm$ 3.2	& 4.4 $\pm$ 1.0		& 0.61 $\pm$ 0.13 \\
Non-$W$		& 58 $\pm$ 12	& 24 $\pm$ 5	        & 2.8 $\pm$ 0.7		& 2.5 $\pm$ 0.8		& 0.20 $\pm$ 0.18 \\
\hline
Bkgd	        & 632 $\pm$ 100	& 279 $\pm$ 43     & 33.6 $\pm$ 5.8	& 11.1 $\pm$ 2.4	& 1.43 $\pm$ 0.59 \\
\ttbar\         & 3.5 $\pm$ 0.4	& 27.9 $\pm$ 3.2   & 52.9 $\pm$ 5.8	& 56.7 $\pm$ 6.1	& 18.2 $\pm$ 2.0 \\
Total		& 635 $\pm$ 100	& 307 $\pm$ 43	   & 86.5 $\pm$ 8.2	& 67.9 $\pm$ 6.6	& 19.6 $\pm$ 2.1 \\
Data & 722 & 346 & 80 & 71 & 23 \\
\end{tabular}
\end{ruledtabular}
\end{table*}
\endgroup

\begingroup
\squeezetable
\begin{table}[htb]
\caption{
Summary of event yields and background (Bkgd) expectations 
sorted by the number of jets in the event,
for events with at least  two $b$-tagged jets,
assuming the measured \ttbar\ cross section of 10.1~pb.
The $H_T$ requirement is released for events with 2~jets. 
The $Z\to\tau\tau$ background is negligible.}

\label{tab:doubletag}
\begin{ruledtabular}
\begin{tabular}{lcccc}
	& $W + 2$ jets	& $W + 3$ jets	& $W + 4$ jets	& $W + \geq$ 5 jets\\ \hline
Dibosons& 0.68 $\pm$ 0.12 & 0.11 $\pm$ 0.02 & 0.041$\pm$ 0.010 & 0.017 $\pm$ 0.005 \\
$t$     & 2.2 $\pm$ 0.5   & 0.53 $\pm$ 0.15 & 0.12 $\pm$ 0.04  & 0.026 $\pm$ 0.010 \\
$W\bbbar$   & 11.3 $\pm$ 3.6  & 1.3   $\pm$ 0.4    & 0.21  $\pm$ 0.08   & 0.010 $\pm$ 0.004 \\
$W\ccbar$   & 0.92 $\pm$ 0.41 & 0.22  $\pm$ 0.11  & 0.07  $\pm$ 0.04   & 0.003 $\pm$ 0.002 \\
$Wc$     & 0.51 $\pm$ 0.15 & 0.05  $\pm$ 0.03  & 0.021 $\pm$ 0.012  & 0.001 $\pm$ 0.001 \\
$W+LF$   & 2.5 $\pm$ 1.3   & 0.02  $\pm$ 0.11  & 0.00  $\pm$ 0.12   & 0.16  $\pm$ 0.23  \\
Non-$W$ & 0.59 $\pm$ 0.59 & 0.28 $\pm$ 0.28  & 0.5    $\pm$ 0.5     & 0.35  $\pm$ 0.35  \\ 
\hline
Bkgd	& 18.7 $\pm$ 4.3	&  2.5 $\pm$ 1.2  &  1.0 $\pm$ 0.8	& 0.6 $\pm$ 0.5 \\
\ttbar\ &  7.5 $\pm$ 1.3& 18.5 $\pm$ 3.2  & 23.7 $\pm$ 4.1	& 7.9 $\pm$ 1.4 \\
Total	& 26.3 $\pm$ 4.5& 21.0 $\pm$ 3.4  & 24.6 $\pm$ 4.2	& 8.4 $\pm$ 1.4 \\
Data	& 30	        & 23	          &  25	                & 6 \\
\end{tabular}
\end{ruledtabular}
\end{table}
\endgroup

\begingroup
\squeezetable
\begin{table}[htb]
\caption{Summary of $b$-tagging information for the algorithms used, where $\epsilon_b$ is the
$b$-tagging efficiency for $b$-quark jets in \ttbar\ events and $\epsilon_b^{\ttbar}$ is 
the per-event efficiency for \ttbar.}
\label{tab:tagsummary}
\begin{ruledtabular}
\begin{tabular}{lcccc}
   		   & \multicolumn{2}{c}{\secvtx} &  {\tt T\secvtx} &  {\tt Jet Probability} \\ 	\hline
$\epsilon_b$ (\%)& \multicolumn{2}{c}{$48 \pm 4$}   & $40 \pm 3$	& $35 \pm 3$ \\		
Mistag rate (\%)& \multicolumn{2}{c}{$1.20 \pm 0.07$}& $0.48 \pm 0.04$	& $1.22\pm 0.08$ \\ \hline		
$b$-tags per event & $\geq 1$	&  $\geq 2$  	&  $\geq 1$	&  $\geq 1$	  	 \\
$\epsilon_b^{\ttbar}$ (\%)&$69 \pm 5$	&$23 \pm 3$	&$60 \pm 3$	& $54 \pm 4$	  	 \\		
Observed $\frac{\mathrm Signal}{\mathrm Background}$	   & 2.8 	& 12.2  & 4.5 & 4.7	 \\		
\end{tabular}
\end{ruledtabular}
\end{table}
\endgroup



\begin{thebibliography}{99}
\bibitem{ref:coords} We use a cylindrical coordinate system about the beam axis 
in which $\theta$ is the polar angle, 
$\phi$ is the azimuthal angle, and $\eta \equiv - \ln \tan(\theta/2)$.  
$E_T \equiv E\ \sin \theta$ and $p_T \equiv p \sin \theta$ 
where $E$ is the energy measured in the calorimeter and $p$ is the 
momentum measured by the spectrometer.  
The missing $E_T$ ($\not\!\! \vec{E}_T$) is defined by $
     \not\!\! \vec{E}_T = - \sum_{i} E_T^i \hat{n}_i$,
where $i =$ calorimeter tower number with $|\eta| < 3.6$
and $\hat{n}_i$ is a unit vector perpendicular to the beam axis and
pointing at the i$^{th}$ calorimeter tower. We also define
$\not\!\!E_T=|{\not\!\!\vec{E}_T}|$.
\bibitem{ref:isolation} We require the calorimeter transverse energy in a 
cone of  $\Delta R \equiv \sqrt{\Delta \eta^2 + \Delta \phi^2} < 0.4$ 
around the lepton (not including the lepton itself) 
divided by the electron (muon) transverse energy (momentum) 
to be less than 0.1.
\end{thebibliography}
\end{document}